\documentclass[twocolumn,amssymb,showpacs,amsmath,nobibnotes,nofootinbib,aps,prb]{revtex4}
\usepackage[dvips]{graphicx}
\usepackage[dvips]{color}
\begin{document}

\title{Investigation of the field-treated magnetic state in Gd$_5$Ge$_4$}
\author{S. Majumdar, S. J. Crowe,  D. McK. Paul, M. R. Lees}
\affiliation{ Department of Physics, University of Warwick, Coventry, CV4 7AL, United
Kingdom.}
\author{V. Hardy}
\affiliation{Laboratoire CRISMAT, UMR 6508, Boulevard du Mar\'{e}chal Juin, 14050 Caen, France.}

\begin{abstract}
At low temperatures, the intermetallic compound Gd$_5$Ge$_4$ shows a sharp field-induced transition into a ferromagnetic state around 25 kOe of applied field.  The material remains in the field treated ferromagnetic state even when the magnetic field is removed. We have investigated the character of this  field-treated state by magnetization and heat capacity measurements. The nature of the  magnetization and the heat capacity are found to be different above and below a charactersitic temperature $T_{irr} \sim$ 25 K.

\end{abstract}
\maketitle

\section{Introduction}
Recently, a stoichiometric intermetallic compound Gd$_5$Ge$_4$ has been reported to show fascinating  magnetic behavior as a function of magnetic field and temperature.~\cite{lev1, lev2} Gd$_5$Ge$_4$ orders antiferromagnetically below $\sim$ 125 K. With the application of an external magnetic field at low temperatures in the zero-field-cooled state, the compound  shows a sharp magnetization jump at a field $\sim$ 25 kOe.~\cite{lev2} The resulting field-treated (FT) magnetic state is ferromagnetic (FM) in nature and it is highly irreversible in the sense that the  compound remains in the FT magnetic state even when the magnetic field is removed. This FT magnetic state has the characteristics of a soft ferromagnet with a very small coercive field~\cite{lev2}. Another unusual  feature of Gd$_5$Ge$_4$ is that different zero field magnetic phases can be created depending upon the previous magnetic field and temperature history of the sample~\cite{tang}. It has been proposed that for certain values of the applied magnetic field and temperature, FM and antiferromagnetic (AFM) magnetic phases can coexist in this compound.~\cite{lev2} A similar situation of intrinsic magnetic phase separation has been observed in mixed valent metal oxides.~\cite{dagotto}

\par
A plausible explanation for the sharp magnetization jump in Gd$_5$Ge$_4$ is that it arises from a martensitic character for the transition involving a lattice distortion.~\cite{lev3} With increasing  magnetic field, the magnetic energy favors the growth of the FM phase, whereas the elastic energy associated with the strains created at the AFM/FM boundaries hinder this process. The magnetization jump corresponds to an avalanche-like growth of the FM component when the driving magnetic field overcomes the energy barriers associated with the boundary strain. Once the  avalanche-like process is complete, the elastic energy  dominates, and the system is locked into the resulting FM state. Interestingly, very similar magnetization jumps have been  observed in  phase separated manganese oxides  with the general formula Pr$_{1-x}$Ca$_x$Mn$_{1-y}$M$_y$O$_3$ ($x \sim$ 0.5, $y \sim$ 0.05, and M is a cation such as Ga or Co).\cite{maignan, mahen, hardy1} The striking similarities in the nature of the magnetization jump and the development of the FT magnetic state in Gd$_5$Ge$_4$ and Pr$_{1-x}$Ca$_x$Mn$_{1-y}$M$_y$O$_3$ suggests that a common martensetic scenario is relevant in both  systems.~\cite{hardy2} 

\par
In order to understand the nature of the FT magnetic state and its relationship with the virgin state, we have performed a thorough investigation of  Gd$_5$Ge$_4$ using magnetization and heat capacity measurements. Our measurements indicate that there is  a characteristic temperature $T_{irr}$, below which a FT ferromagnetic state can be stabilized within this material.

\begin{figure}[t]
\centering
\includegraphics[width = 7.5 cm]{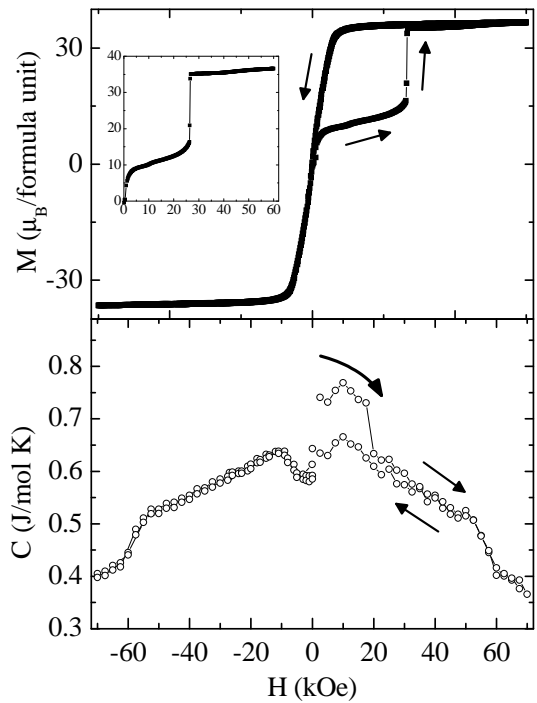}
\caption{The top panel shows a six quadrant $M$ versus $H$ loop of Gd$_5$Ge$_4$ measured at 2 K. The inset in the top panel shows an enlarged view of the virgin magnetization loop. The bottom panel shows a four quadrant $C$ versus $H$ data recorded at 2 K.}
\label{6QMH}
\end{figure} 

\section{Experimental details}
Polycrystalline Gd$_{5}$Ge$_{4}$ samples were prepared by arc melting elemental  Gd (99.9 wt\% purity) and Ge (99.99 wt\% purity) in the stoichiometric ratio under a high-purity argon atmosphere. The Gd/Ge ratio was checked by Energy Dispersive Spectroscopy (EDS) to be equal to the nominal composition to within the accuracy of this technique (about 2\% of the elemental ratios). No impurity phases were detected by x-ray powder diffraction, which showed that the system has an orthorhombic structure at room temperature ($Pnma$ $\ $space group) with lattice parameters [$a=7.68(1) $ \AA $,$ $b=$ $14.80(1)$ \AA\ and $c=7.77(1)$ \AA ] in line with the literature.\cite{lev1} Magnetic measurements were carried out using a Quantum Design Superconducting Quantum Interference Device (SQUID) magnetometer and an Oxford Instruments Vibrating Sample Magnetometer (VSM). The  heat capacity ($C$)  measurements were performed  by a relaxation method using a Quantum Design Physical Properties Measurement System (PPMS). We note that the transition field in Gd$_5$Ge$_4$ is not a true critical field~\cite{hardy2}, since it varies  with the field sweep rate and also to some extent on the previous magnetic history (training effect) of the sample. The field sweep rate in the various apparatus used for this study are different, and it is likely that the precise value of the transition field may be different.

\section{Results}
A six quadrant magnetization versus field  loop for Gd$_{5}$Ge$_{4}$ taken at 2 K is shown in figure~\ref{6QMH}. The virgin curve (obtained after zero-field cooling from 300 K) shows the magnetization jump around 25 kOe followed by FM-like  behavior characteristic of this material. The field decreasing part of the magnetization loop follows a different path, and it is clear from the subsequent field cycles that once the FM phase is established in the sample it is quite stable; the sample retains this FM character even when the magnetic field is set to zero. The FT magnetic state shows  magnetization behavior which is typical of a very soft ferromagnet. The value of the coercive field at 2 K is found to be 30 ($\pm$ 5) Oe. The spin stiffness constant ($D$) in the FM state is related to the $T$-dependence of the saturated magnetization (see equation (2) of reference~\cite{roy}). In the present case,(data not shown) we have found the value of $D$ in the FT magnetic state to be 14($\pm$ 3) meV\AA$^2$. The inset of figure 1 shows  an enlarged view of the virgin curve. The overall behavior displayed here is in line with the features reported by Levin et {\it et al.}\cite{lev2} Our sample, however, exhibits a sizable FM component at low fields, as revealed by the shape of the virgin magnetization curve and its $M(H)$ behavior (figure~\ref{6QMH}) is closer to that described in reference~\cite{lev2} when the sample was cooled in a field of 1.2 T to assist the onset of the FM component. The two samples may differ on a  microstructural level, which may influence the ability of the samples to accommodate the strains associated with the martensitic FM phase.\cite{nis, podz}

Figure~\ref{6QMH} (lower panel) shows a four quadrant heat capacity versus magnetic field data set recorded at $T$ = 2 K after the sample was zero-field-cooled (ZFC) from 300 K. The virgin, field increasing  branch, shows a sharp jump around 20 kOe which corresponds to the AFM-FM martensitic transition seen in the magnetization data. The small difference in the value of the field at which the transition occurs in the $M(H)$ and $C(H)$ measurements may be due to the different field sweep rates used. In agreement with the magnetization data, the subsequent branches of  the $C$ versus $H$ data are reversible and deviate markedly from the initial field increasing leg  below $H$ = 20 kOe.  That is, the sharp jump followed by the reversible behavior in the magnetization data is also reflected in the thermodynamic response of the system.

\begin{figure}[t]
\centering
\includegraphics[width = 8 cm]{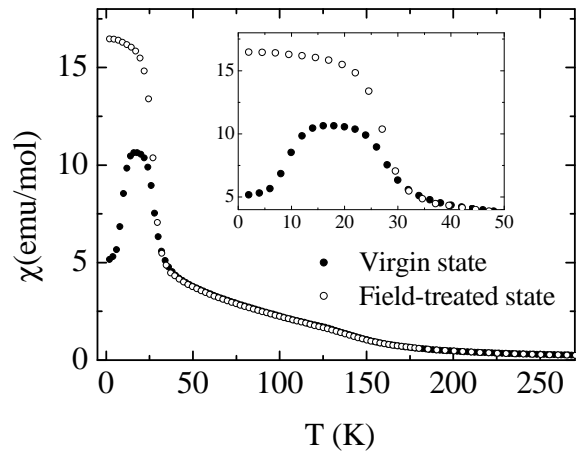}
\caption{ $M$ versus $T$ data of Gd$_5$Ge$_4$ in an applied field of 12 kOe recorded both in the virgin state and in the field treated state. The inset shows an enlarged view  of the low temperature (below 50 K) behavior.}
\label{MT}
\end{figure}

\begin{figure}[t]
\centering
\includegraphics[width = 7.5 cm]{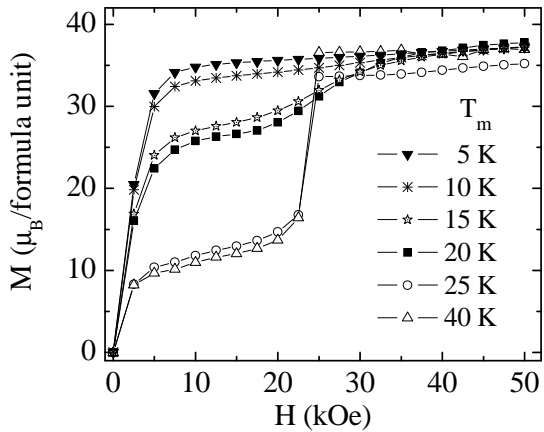}
\caption{$M$ versus $H$ data recorded at 2 K after the field-treated sample has been heated to the temperature $T_m$  and subsequently cooled to 2 K in zero field.}
\label{Tm}
\end{figure}

We have performed  magnetization  measurements on the sample in  both  the FT and virgin magnetic states in order to investigate the nature of the magnetic state at low temperatures. The FT magnetic state was prepared by the application of a magnetic field of 50 kOe at 2 K in the ZFC state for 15 minutes followed by a subsequent removal of the field. Figure~\ref{MT} shows the the magnetic susceptibility ($\chi$) versus temperature data recorded on the virgin state (ZFC heating) and the FT magnetic state (heating) in an applied magnetic field of 12 kOe. The high temperature (above $\sim$ 25 K) $\chi(T)$ data for the virgin state and the FT magnetic state are identical, and both show a kink around 127 K in the $\chi(T)$ data and a sharp peak at 123 K in the $\frac{d\chi(T)}{dT}$ data. On heating from low temperatures, the ZFC data in the virgin state show a pronounced rise in the susceptibility (at $T \sim$ 4 K), followed by a plateau and  then a sharp drop above 22 K. This indicates the existence of a complex magnetic behavior in the system~\cite{lev2} with a predominantly  AFM ground state. The $\chi(T)$ data in the FT magnetic state and in the  virgin state deviate from one another below an irreversibility temperature $T_{irr}$ = 25 K.  At low temperatures the magnitude of $\chi(T)$ in the FT state is larger than $\chi(T)$ in the virgin state  and  saturates below $\sim$ 15 K. This behavior is consistent with the  FM  nature of the FT magnetic state.

In order to better understand the significance of $T_{irr}$, we have collected  $M$ versus $H$ data at 2 K  after the following sequence of field and temperature treatments. The sample was first ZFC to 2 K from 300 K and a 50 kOe field was applied in order to create the FM state. The field was then removed and the sample was heated to a temperature $T_m$ and then ZFC down to 2 K. The $M$ versus $H$ data were then collected on the sample.
This cycle was repeated for different values of $T_m$. It is clear from the data (figure~\ref{Tm}) that for $T_m$ $<$ 25 K, the system remains in  the FM state, while for $T_m$ $\geq$ 25 K, the system return to the virgin state and a jump in the $M$ versus $H$ data is once again observed.  The value of $T_m$ for which this  change in behavior is observed  coincides with $T_{irr}$.

\par
\begin{figure}[t]
\centering
\includegraphics[width = 7.5 cm]{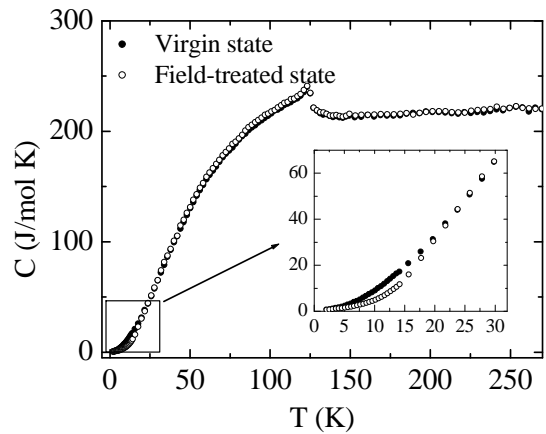}
\caption{Zero field $C$ versus $T$ data recorded on both the virgin and field-treated states of Gd$_5$Ge$_4$. The inset shows the low temperature data in more detail.}
\label{CT}
\end{figure}

\begin{figure}[t]
\centering
\includegraphics[width = 7.5 cm]{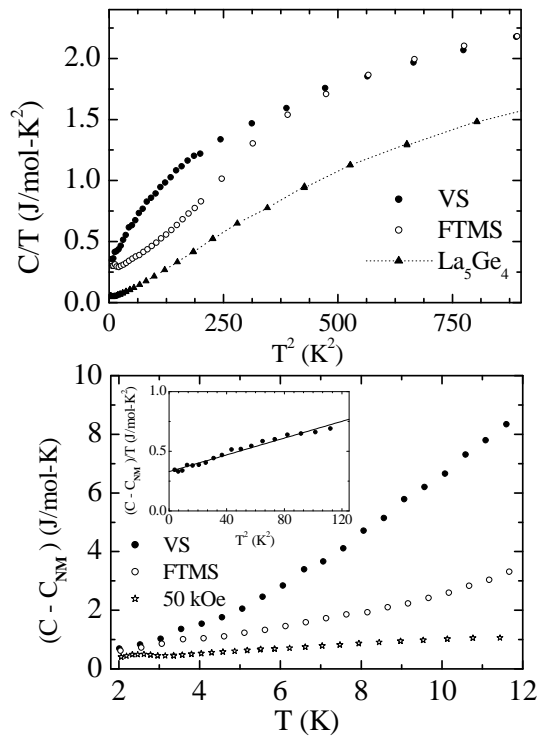}
\caption{The top panel shows the $C/T$ versus $T^2$ data for the virgin state  and the field-treated magnetic state  of Gd$_5$Ge$_4$, along with the data of the non-magnetic counterpart La$_5$Ge$_4$. The bottom panel shows  $\left(C - C_{NM}\right)$ versus $T$ data of Gd$_5$Ge$_4$ in the virgin state, FT magnetic state and at an applied field of 50 kOe, where $C_{NM}$ represent the non-magnetic contribution obtained from the $C$ data of La$_5$Ge$_4$. The inset in the bottom panel shows the  $\left(C - C_{NM}\right)/T$ versus $T^2$ data in the virgin state along with a linear fit to the data (solid line).}
\label{cmag}
\end{figure}

 
We have also measured the heat capacity  of Gd$_5$Ge$_4$  as a function of temperature in both the  virgin and the FT magnetic state (see figures~\ref{CT},~\ref{cmag}).  The zero field $C(T)$ data were collected while heating the sample from 2 K to 300 K. A  peak is observed in the $C(T)$ data at $T_N$ = 123 K. The data collected in the FT magnetic state and virgin state are identical above 25 K, while a clear deviation between the two data sets is observed at low temperatures. Below 25 K, $C(T)$ for the FT state is lower in magnitude than $C(T)$ for the virgin state at the same temperatures, indicating a difference in the magnetic character of the two states. The upper panel of figure~\ref{cmag} shows the $C(T)/T$ versus $T^2$ data below 30 K for the FT magnetic state, virgin state of Gd$_5$Ge$_4$ along with  data for the nonmagnetic isostructural compound La$_5$Ge$_4$. The $C(T)/T$ data (both in FT magnetic state and virgin state) are larger  for the Gd compound than the La counterpart, which is  due to an additional magnetic contribution to the heat capacity. 

\par
The temperature dependence of the magnetic contribution to the heat capacity can provide us with important information about the nature of the magnetic order in a system. We have analyzed the data  by assuming that $C$ comprises three contributions, namely  a phonon contribution from the lattice ($C_{latt}$), an  electronic contribution ($C_{el}$) and the magnetic contribution $C_{mag}$. It is always difficult to obtain a correct estimate for the value of the two nonmagnetic contributions ($C_{NM} = C_{latt} + C_{el}$) over a large temperature range.  Here, we have used the $C(T)$ data of La$_5$Ge$_4$ to estimate $C_{NM}$ for Gd$_5$Ge$_4$ using the method of Bouvier {\it et al.}~\cite{bouvier}, which is a good approximation at least at low temperatures, where $C_{latt}$ is small. The lower panel in figure 4 shows the heat capacity  as a function of temperature with $C_{NM}$ subtracted  from $C$.  For the virgin state,  $\left(C-C_{NM}\right)/T$ versus $T^2$ data shows a linear behavior (inset in the lower panel of figure 4), as expected for an AFM contribution to the heat capacity with a $T^3$ dependence. However, the finite value of $\left(C-C_{NM}\right)/T$ obtained from the intercept of a  linear fit to the data, indicates that $\left(C-C_{NM}\right)$ also has a $\gamma T$ contribution, where $\gamma$ is 0.07(1) JK$^{-2}$ per Gd atom. A linear contribution to the magnetic heat capacity can arise from some intrinsic disorder in the system, which is not unexpected for a phase-separated system such as Gd$_5$Ge$_4$. For example, a linear term in the magnetic heat capacity is also reported for manganites~\cite{smo,hardy3}, although here our $\gamma$ value is one order of magnitude higher than those reported in references~\cite{smo,hardy3}, whilst a  Gd$_5$Ge$_4$ sample with a smaller FM component in the ZFC state was reported to have $\gamma$ = 0.013 JK$^{-2}$ per Gd atom~\cite{lev1} .

\par
The $\left(C-C_{NM}\right)/T$ versus $T$ behavior in the FT magnetic state also shows a large value of $\gamma$ at low temperatures. For a ferromagnetic state one may expect a $T^{3/2}$ behavior for the magnetic contribution to the  heat capacity~\cite{gopal}. The  data for the  FT magnetic state does not follow a simple $T^{3/2}$ law, particularly below
5 K, where there is  a broad peak around 3.5 K. Nevertheless, a  fit to the $\left(C-C_{NM}\right)$ data in the temperature range 5-10 K using the expression  $\gamma T + AT^{3/2}$  gives us a coefficient for  the $T^{3/2}$ term of  $A$ = 0.066 ($\pm$ .005). The coefficient $A$ is related to the spin stiffness constant $D$~\cite{kittel}, and we obtained a value of $D$ equal to  18 mev \AA$^2$ ($\pm$ 2). This value is consistent with the value of $D$ (= 14 mev \AA$^2$($\pm$ 2)) obtained from the $M(T)$ data.    

\par
We have also measured the heat capacity in an applied magnetic field of 50 kOe after zero-field-cooling from  room temperature. In 50 kOe, $\left(C-C_{NM}\right)$ is suppressed to a value even lower than that of the FT magnetic state state (figure~\ref{cmag}, bottom panel). The $\left(C-C_{NM}\right)$ shows a complex behavior at low temperatures with an peak below 3 K, possibly a weaker version of which is also present in the FT state data. A  peak  is also observed at 50 K, as previously reported  by Tang {\it et al.}~\cite{tang}. This has been attributed  to a first order phase transition from the low temperature FM state to a high temperature AFM phase.

\begin{figure}[t]
\centering
\includegraphics[width = 7.5 cm]{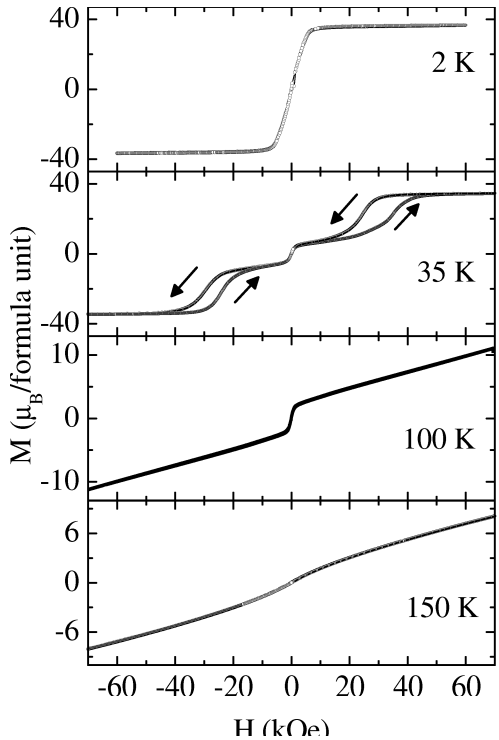}
\caption{Four quadrant magnetization versus field loops recorded at 2 K, 35 K, 100 K  and 150 K. Whilst the  2 K data has been obtained in the field-treated magnetic state, the $M(H)$ data shown for the field increasing and decreasing legs at the other temperatures shown were collected from a virgin state, as above 25 K, the field-treated state and the virgin state are indistinguishable.}
\label{MH}
\end{figure}

\begin{figure}[ht]
\centering
\includegraphics[width = 7.5 cm]{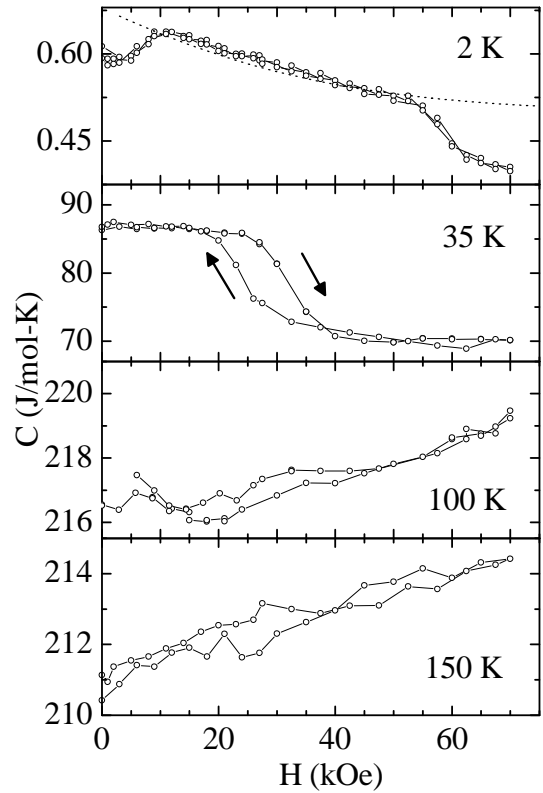}
\caption{Two quadrant heat capacity versus field loops recorded at 2 K , 35 K, 100 K and 150 K. The  2 K data was obtained in the field-treated magnetic state. The $C(H)$ data shown for the field increasing and decreasing legs at  the other  temperatures shown were obtained from a virgin state, as above 25 K, the field-treated state and the virgin state are  indistinguishable. The dotted line in the top panel (2 K data) shows an estimate for $C(H)$ behavior calculated using a spin wave model (see text for details).}
\label{CH}
\end{figure}

We have measured the magnetization and the heat capacity as a function of the applied magnetic field at different temperatures above and below $T_{irr}$. Figure~\ref{MH} shows the $M$ versus $H$ data recorded at 2 K, 35 K, 100 K, and 150 K, while figure~\ref{CH} shows the $C$ versus $H$ data at the same temperatures. The data at 2 K for both the $M$ and $C$ measurements are recorded in the FT magnetic state. A comparison of the $M(H)$ data at 2 K and 35 K highlights the difference in the nature of the magnetic state of the system at these two temperatures. A fully saturated FM state can also be obtained at 35 K (above $T_{irr}$). After the first leg of the field cycle, however, the sample continues to exhibit a step in the data and a hysteresis between 10 to 50 kOe in both positive and negative $H$. Below 10 kOe the data coincide with the virgin curve.  This indicates that the field induced FM state at 35 K only exists at high fields and that the sample returns to a phase separated AFM-FM  state as soon as the field is removed. In contrast, at 2 K, once the field induced FM state is established, it  exists even at low field. The $M(H)$ data recorded at 100 K  show no sign of a field induced FM state as is  evident from the absence of saturation at the highest field. Apart from an initial rise, the $M(H)$  data shows a linear behavior, typical of an AFM state. Above  $T_N$ (= 123 K), the $M$ versus $H$ data  exhibit a slight positive curvature with respect to the field axis (see data at 150 K), that may be due to spin fluctuations which are a precursor to the long range AFM order.

\par
The $C$ versus $H$ data are consistent with the $M(H)$ measurements (figure~\ref{CH}). The FT magnetic state at 2 K is non-hysteretic, signaling that the sample has locked into the field induced soft FM state. The 2 K $C(H)$ data shows some complex features at low applied field (below 10 kOe) and also at high field above 50 kOe. The low field features in the $C(H)$ curves correspond to the sharp rise in $M$ before saturation is observed; complex domain movements are involved in this process, which cannot be accounted for by any simple theory. (Note, that these low field features are also present in the virgin state (see figure~\ref{6QMH}) where there is also a decrease of 0.1 J/mol-K ($\sim$ 14\% of the zero field value) in the $C$ versus $H$ curve  at a field of $\sim$ 20 kOe.) In  the  region of 10-50 kOe, $C(H)$ decreases smoothly, as expected for a FM system~\cite{roy, roy1}, i.e., $\left(\frac{\partial C}{\partial H}\right)_T <$ 0. The field dependence of $C$ is consistent with the predicted $C(H)$ behavior using a simple model for ferromagenetic spin-waves (equation 1 of reference~\cite{roy}) with $D$ = 14 meV\AA$^2$.  The sharper fall in  $C(H)$ above 50 kOe corresponds to the region in the $M(H)$ data where the system  attains full saturation. The $C(H)$ data at 35 K contain hysteresis in the same magnetic field range as the  $M(H)$ data. It  shows  a large decrease ($\sim$ 16 J/mol K which is 20\% of the zero field value) with increasing field. Either side of this jump,  the $C(H)$ data are found to be almost field independent. This may be a reflection of the combined field response of the  FM and AFM components in the system. The $C$ versus $H$ data at 100 K and 150 K increase slightly with field ($\sim$ 4 J/mol K which is only 2\% of the zero field value), and reflects the existence of  AFM correlations at these temperatures, where we expect $\left(\frac{\partial C}{\partial H}\right)_T >$ 0.

\section{Discussion}

The present magnetization and heat capacity investigations on Gd$_5$Ge$_4$ address the nature of the field-treated magnetic state. The FT magnetic state is established via the application, at low temperatures, of a magnetic field of  $H \sim$ 25 kOe after zero field cooling. Our measurements indicate that the FT magnetic state is removed if the sample is heated above $T_{irr}$ = 25 K. We believe that  $T_{irr}$ is a characteristic temperature for Gd$_5$Ge$_4$, in the sense that it is the  temperature below which an FT magnetic state can be stabilized in this system. 

\par
Tang {\it et al.} observed a change in the reversible character of the  $M(H)$ data at 21.3 K~\cite{tang}. This temperature corresponds closely to $T_{irr}$ observed in our $M(T)$ and $C(T)$ data (see figure 3) measured for virgin state and FT magnetic state. Changes in the nature of the $M(H)$ and the $C(H)$ data above and below $T_{irr}$ are also observed. This implies that not only $M$, but the magnetic contribution to $C$ also undergoes a change in character across $T_{irr}$.

\par
The large change in $C(H)$ at 35 K is consistent with the observation of giant magneto-caloric effect (MCE) in this material. At 35 K, Gd$_5$Ge$_4$ shows a change in the magnetic entropy -$\Delta S_{mag} \sim$ 25 J/mol-K for a magnetic field change from 0 to 50 kOe.~\cite{pecharskymce} Recent X-ray diffraction studies on Gd$_5$Ge$_4$ show a field induced structural transition (FIST) in this material,~\cite{pecharsky2} which corresponds to the jump in the magnetization and heat capacity. It is evident that  there is a strong magneto-structural coupling in this compound and the large MCE observed here is related to the FIST.  The large irreversible change in the $C$ versus $H$ data seen at 2 K (see figure~\ref{6QMH}) and the reversible change seen at 35 K (figure~\ref{CH}) are clearly due to the associated structural transition triggered by the magnetic field. The magnitude of the change at 35 K is much larger than that seen at 2 K due to the larger lattice contribution at 35 K. The nature of the FIST is found to be different at 6.1 K and 25 K, in the sense that the structural change is reversible at 25 K, whereas  at 6.1 K,  once the FIST occurs the sample remains in the transformed state upon removal of the field. This behavior matches well with the data presented here where we have identified a characteristic temperature $T_{irr} \sim$ 25 K for this system.

\par
We acknowledge the financial support of the EPSRC (UK) for this project.

\end{document}